\documentclass[twocolumn,twoside,slac_two]{revtex4}
\usepackage{graphicx}
\usepackage{fancyhdr}
\usepackage{hyperref}
\hypersetup{
    colorlinks=true,
    urlcolor=magenta,
}
\begin{document}

\title{High-energy electron bursts in the inner Earth magnetosphere caused by precipitation from radiation belt}


\author{T.~Zharaspayev, S.~Aleksandrin, A.~M.~Galper, S.~Koldashov}
\affiliation{NRNU MEPhI, Kashirskoe highway 31, 115409 Moscow, Russia}

\begin{abstract}
Orbital experiment ARINA on the board of Russian satellite Resurs-DK1 launched in 2006 developed to study charged particle flux (electrons $E \sim 3 - 30 MeV$, protons $E \sim  30 - 100 MeV$) in near-Earth space, especially high-energy electron precipitation from the inner radiation belt caused by various geophysical and solar-magnetospheric phenomena. Precipitated electrons under certain conditions (energy, LB-coordinate) drifts around the Earth and can be detected as fast increase in count rate of satellite spectrometer (so called bursts).  High-energy electron bursts can be caused by local geophysical phenomena (like earthquakes or thunderstorms). Such bursts have distinct features in their measured energy-time distribution. These features contains information about initial location of electron precipitation. In previous works, particle precipitation region searching method is described, the main idea of the method is to use numerical model of electron movement in magnetosphere to find longitudinal distance between region of precipitation and burst registration location on the board of satellite, and with knowledge of L-coordinate define precipitation region borders. Major problem of this type of analysis is the high number of background electrons (atmospheric albedo). Several methods (linear, robust regression) were used previously to minimize number of background particles involved in analysis. In this report, the new ensemble method was developed, it uses the combining results from several methods in dependence of burst registration conditions. Ensemble method was tested on simulation and experimental data. Numerical simulation of local particles precipitations based on well-known equations of relativistic particle movement in Earth magnetosphere. In experimental data analysis, the results from ARINA experiment for $10$ years was used. Several results based on burst experimental data are shown. Ensemble method shows better results than any single method alone.\end{abstract}

\maketitle

\thispagestyle{fancy}


\section{Introduction}

The Earth's magnetic field at a distance of up to three of its radius has a dipole structure, after which the magnetic field begins to be affected by the solar wind.
	
Charged particles can be caught in this field and begin to participate in the three types of movement - rotation around the magnetic field lines, the reflection between the mirror points and longitudinal drift. The area in which these particles are trapped for a long time is called the radiation belt.

Various geophysical phenomena such as lightnings or earthquakes due to the emission of low-frequency electromagnetic radiation propagating upward, can lead to a disturbance of the trajectory of the particles which leads to lowering of their mirror points. Such precipitated particles may be registered in the spacecraft crossing the magnetic drift shell in the form of a rapid increase in the rate of the detector counting, this phenomenon referred to as bursts of particles. \cite[and references therein]{burstcite}

\section{Satellite experiment ARINA}

ARINA experiment on board the spacecraft Resurs-DK1 \cite{arinaexp} launched in 2006 and was in operation until earlier this year. The experiment was aimed to study the phenomenon of particle bursts. The equipment of the experiment is able to record electrons with energies ranging from 3 to 30 MeV with an accuracy of 15\%

\begin{table}
	\begin{center}
		\caption{Arina experiment specifications}	
		\begin{tabular}{ | l | c | c | }
			
			\hline
			\centering
			
			\textbf{\emph{Characteristics}} 			& 		&  Value \\
			\hline
			\centering
			\textbf{Geometry factor}			&		& $ 10 \text{cm}^{2}\text{sr} $	\\
			\hline
			\centering
			\textbf{Apperture}				&		& $ \pm 30 \text{degree} $	\\
			\hline
			\centering
			\textbf{Energy}		& protons	& $ 30 - 100 \text{MeV} $	\\
			& electrons	& $ 3 - 30 \text{MeV} $		\\
			\hline
			\centering
			\textbf{Energy resolution}		& protons	& $ 10\% $			\\
			& electrons	& $ 15\% $			\\
			\hline
			\centering
			\textbf{Time resolution}			& 		& $ 100 \text{ns} $		\\
			\hline
			\centering
			\textbf{Weight}					&		& $ 8,6 \text{kg} $		\\
			\hline
			\centering
			\textbf{Power}			&		& $ 13,5 \text{W} $		\\
			
			\hline
			
		\end{tabular}
		\label{aritab}
	\end{center}
\end{table}

\section{Physical model of longitude electron drift}

The registration process of precipitating cloud of particles by the spacecraft looks simple. A cloud of simultaneously precipirated from the radiation belt electrons with different energies, begins to drift around the Earth to the east within its drift shell, so it can be registered on board the satellite anywhere in the near-Earth space at any longitude when satellite crosses this drift shell.

At the same time, due to the fact that the velocity of the particle longitudinal drift depends on their energy, the electron cloud begins to stretch the greater the longer it is moves around the Earth. Due to this effect, the electron bursts carry information about the location of the precipitation from the radiation belt.
	
\begin{equation}
\tau = \frac{88(1+E/E_{0})}{2+E/E_{0}}\frac{K}{LE}
\end{equation}

\begin{equation}
K = 1.25-0.25\cos^{2}\lambda_{m}
\end{equation}
	
$\tau$ -- Time of one revolution of the particles around the Earth in minutes.

$\lambda_{m}$  -- geomagnetic latitude of reflection point.
	
$E$  -- particle energy.
	
$E_{0}$ -- rest mass of the particle.
	
\begin{figure}
	\centering
	\includegraphics[width=0.5\textwidth]{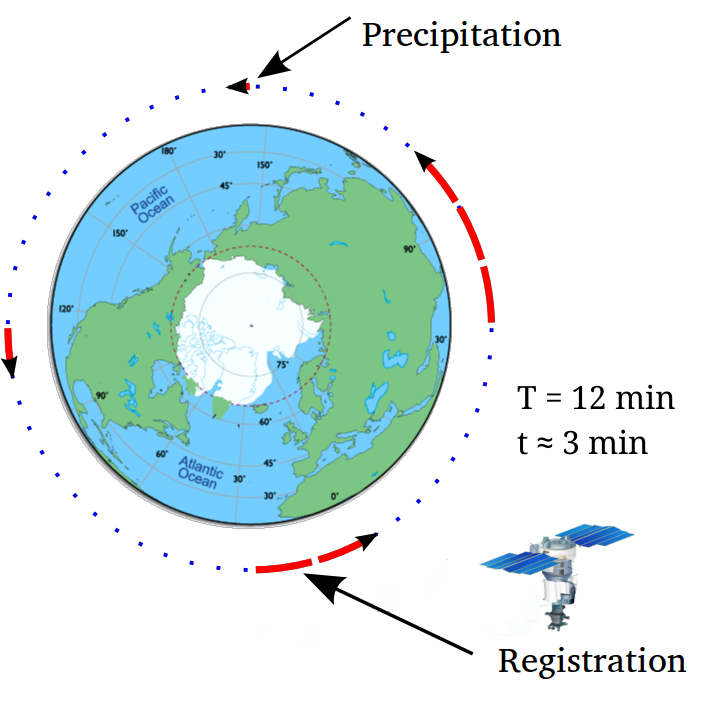}
	\caption{
		Model of the longitudinal electron drift from their place of precipitation from the radiation belt to point of registration on board the satellite. ($t$ -- maximum time satellite crossing disturbed $L$-shell, $T$ -- required crossing time to register whole electron cloud after 180 degrees ) }
	\label{figsatmove}	
\end{figure}
	
\begin{figure}
	\centering
	\includegraphics[width=0.5\textwidth]{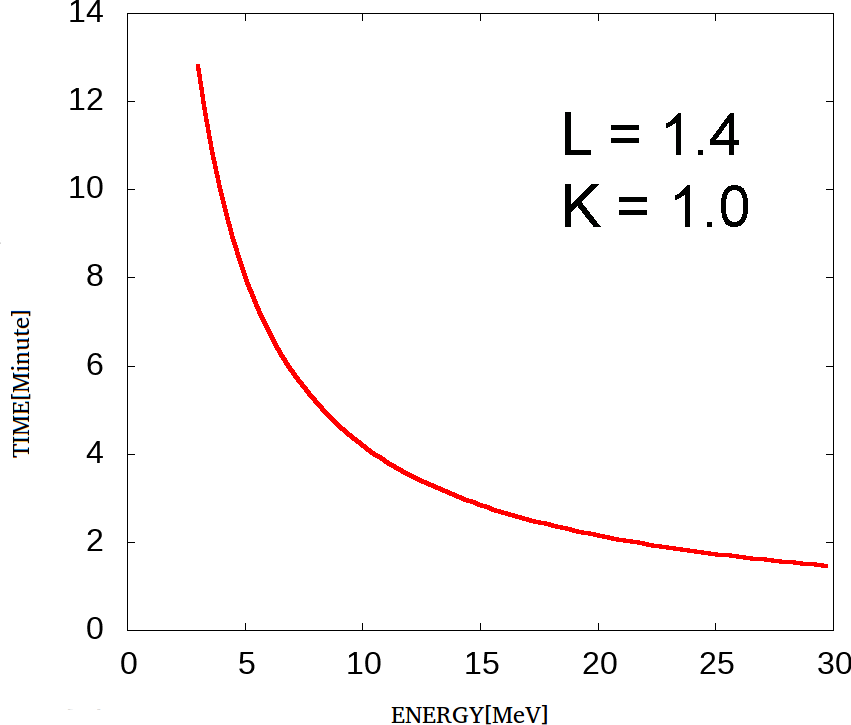}	
	\caption{Expected time curve of one revolution around the Earth of the particle of a given energy}
	\label{fig:expectline}
\end{figure}

\section{Determination of particles cloud precipitation location}

We analyzed a number of methods\cite[and references therein]{icrc2015} for determining the precipitation location of particle clouds from the radiation belt on the basis of time and energy burst electron characteristics. Least-squares method, commonly used for solving problems of this type showed itself deficient in the numerical simulation data, so a comparative analysis of other methods was carried out.
	
\begin{itemize}
	\item Least-squares method \\
	\begin{equation}
	R_{LS} = \sum{ \frac{ (\overline{T_e}-T(\overline{E_e}, t_o, S_d))^2}{\overline{\sigma_e^2}}} \rightarrow  { \min_{R_{LS}}}			
	\end{equation}
	\item Robust regression method \\
	\begin{equation}
	R_{RR} = \sum{ \overline{w} \cdot \frac{ (\overline{T_e}-T(\overline{E_e}, t_o, S_d))^2}{\overline{\sigma_e^2}}} \rightarrow  { \min_{R_{RR}}}		
	\end{equation}
	\item Bootstrap
	\item Committee method \cite{zhou} \\
	\begin{equation}
	S_d = \frac{1}{N_{model}}\sum_{i = 0}^{N_{model}}{W_i \cdot S^i_d}
	\end{equation}
\end{itemize}
	
On their basis the algorithm to determine the geographical coordinates of precipitation location was developed and implemented the software.
	
\begin{itemize}
	\item Null-approximation -- search of burst using time profile data
	\item First approximation -- Least-squares method
	\item Second approximation -- (2 criteria, 2 models) Robust regression
	\item Result -- Committee method	
\end{itemize}

\section{The technique of numerical simulation}

We carried out numerical simulation of burst electrons and fluxes of albedo electron with defined range of energies and power law spectrum. The detector characteristics also took into account.

For the analysis of the data graphs of the time of the particle registration dependency on its energy was used, for brevity called the $ET$-diagrams.		

\begin{figure}
	\centering
	\includegraphics[width=0.49\textwidth]{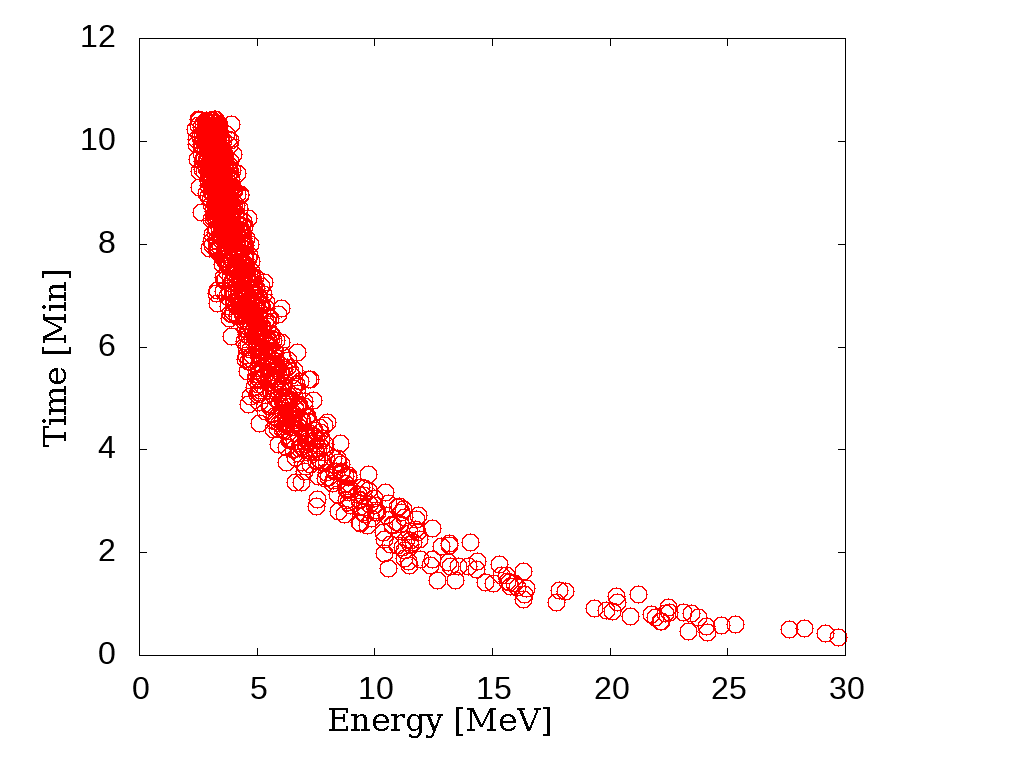}
	\caption{Burst of electrons in the case when the spacecraft was able to observe all of particles of the cloud withoth registering background particles, taking into account the characteristics of the detector. [The result of numerical simulation]}	
	\label{modideal}
\end{figure}

Figure \ref{modideal} shows expected dependency in ideal conditions, which can't be achieved in real environment due to limited time of satellite passing through the disturbed $L$-shell as shown in fig. \ref{figsatmove}.
	
\begin{figure}
	\centering
	\includegraphics[width=0.49\textwidth]{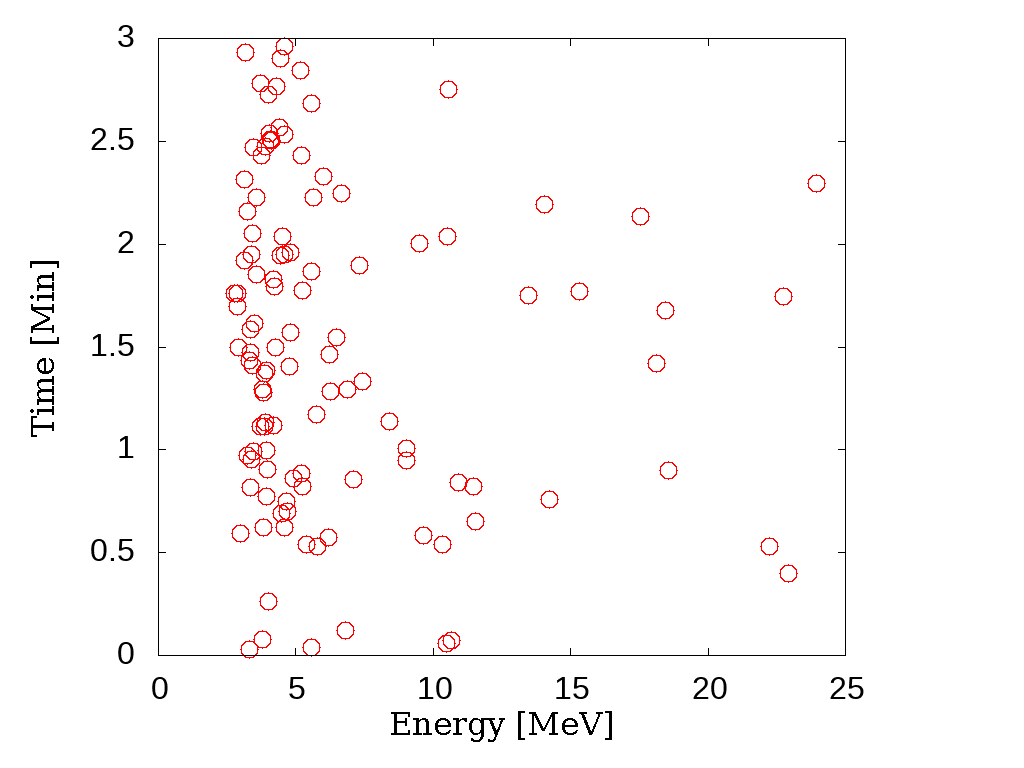}
	\caption{Burst of electrons in conditions close to experimental. (Due to the motion of the satellite, it is possible to observe only a part of the burst, at the same time with the burst electrons albedo particles are recorded, the number of which may be at the level of the number of burst particles) [Result of numerical modeling]}		
	\label{fig:real}
\end{figure}
	
Figure \ref{fig:real} shows that good dependence corresponding to the expected curve shown in fig. \ref{fig:expectline} is difficult to observe.

\section{Time profiles of electron bursts}

Due to the fact that the cloud of precipitating electrons have energy spectrum obeys a power distribution, registered burst of particles will have a distinctive saw-like appearance. Registering the beginning of burst will have a growing profile with a sharp precipice, at the moment, as the spacecraft leaves the disturbed drift shell, while the registration end of the burst is characterized by first rise in count rates and then fall, as a cloud of electrons will continue its movement. Thus, the beginning and end of the burst will have a radically different time profiles.
	
Furthermore, low-energy burst electrons registered with albedo electrons are not distinguishable on $ET$-diagram from the background at all, while in the registration start the burst, their separation can be clearly seen.
	
\section{Burst 24 august 2009}

\begin{figure}
	\centering
	\includegraphics[width=0.5\textwidth]{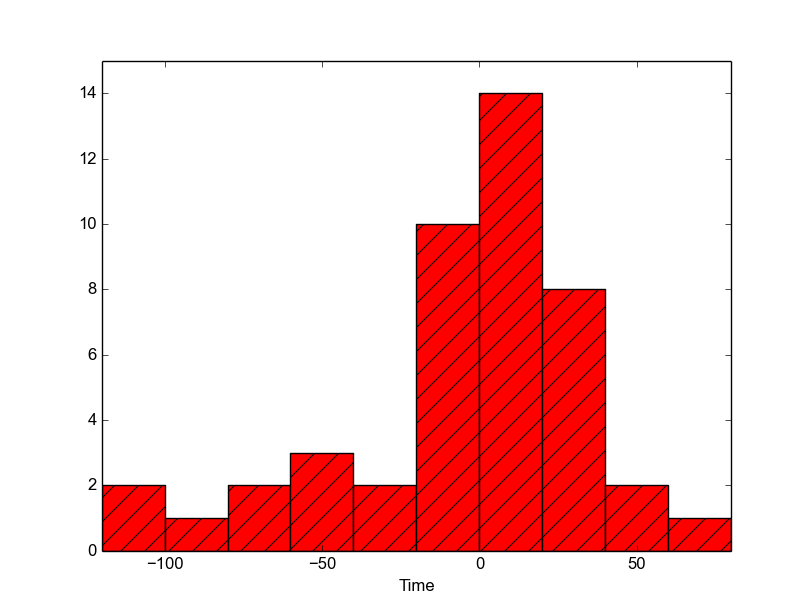}
	\caption{Time profile of burst registered on 27 august 2009}
	\label{tmview}
\end{figure}

\begin{figure}
	\centering
	\includegraphics[width=0.5\textwidth]{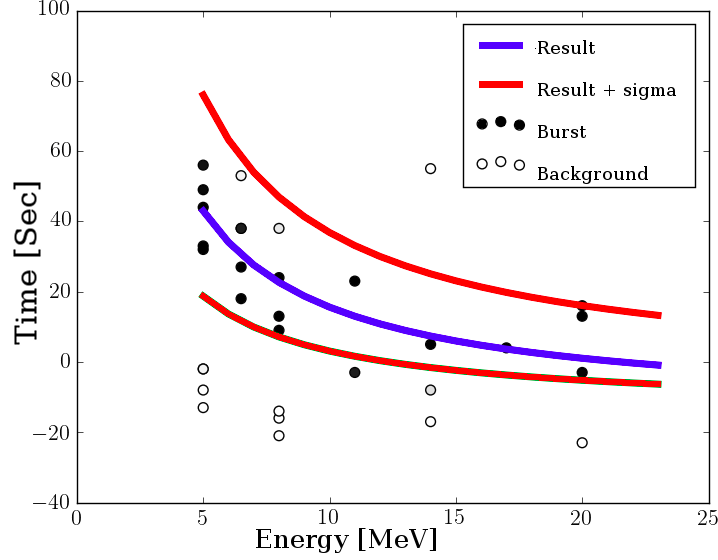}	
	\caption{$ET$-diagram of burst registered on 27 august 2009}
	\label{etview}
\end{figure}

\begin{figure}
	\centering
	\includegraphics[width=0.5\textwidth]{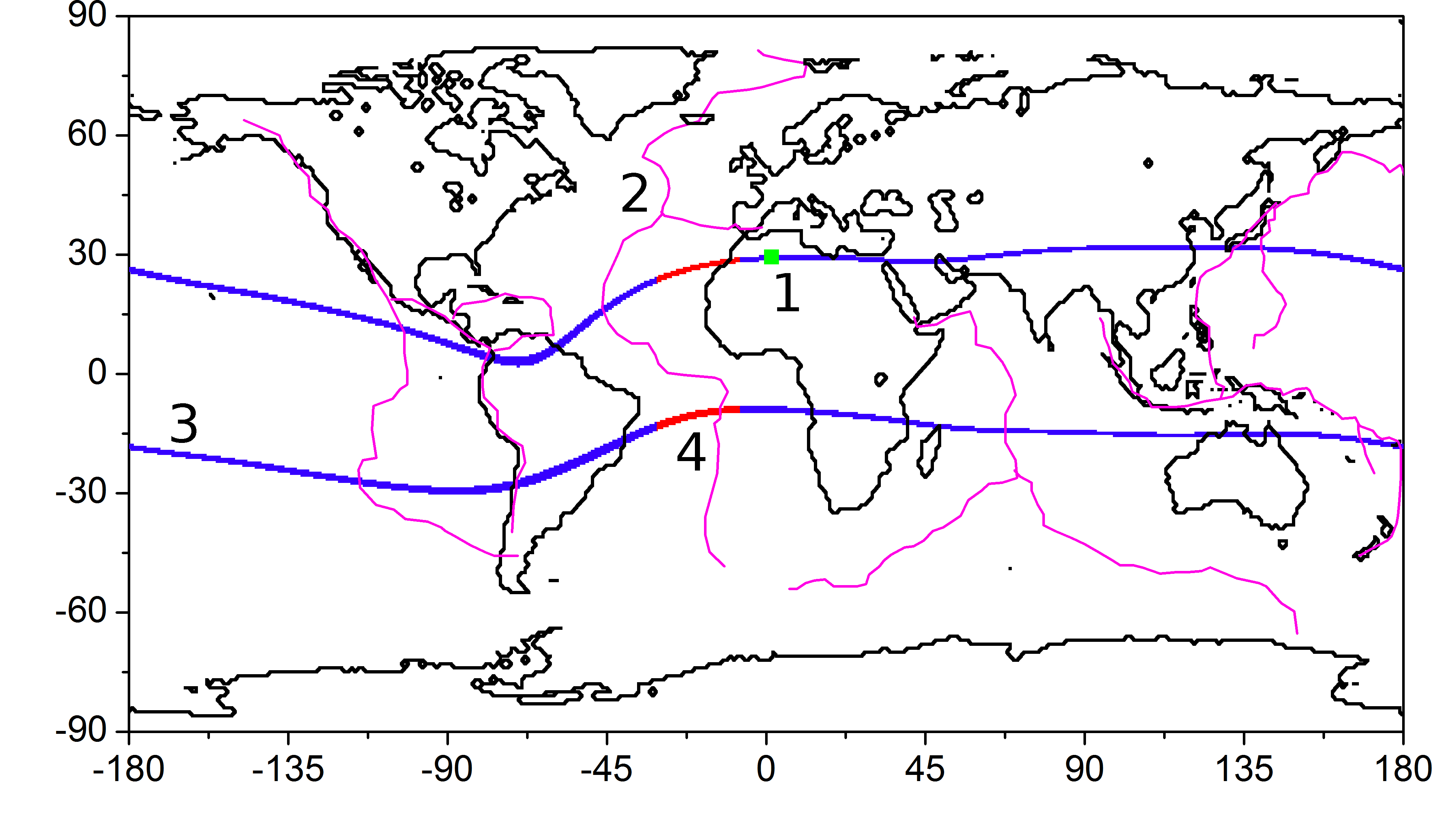}	
	\caption{World map. 1 -- burst registration location, 2 -- Earth's tectonic plates, 3 -- $L$-shell 1.21, 4 -- particle precipitation area}
	\label{mapview}
\end{figure}

Using the developed algorithm, the analysis was conducted on a electron burst selected according to the criteria discussed in this paper, registered at 24 August 2009 (fig. \ref{tmview}, and fig. \ref{etview}). The burst was registered over North Africa in the low $L$-shell 1.21, the burst particles have high energy. Designated particle precipitation area is over the Atlantic Ocean and includes part of the Mid-Atlantic Ridge (fig. \ref{mapview}).

\section{Conclusion}

he aim of this study was to determine the geographic location of the electron precipitation zone from the inner radiation belt under the influence of local geophysical phenomena. To determine the latitude of precipitation area data on $L$-shell was used, while for determining longitude the developed in this study algorithm based on the method of least squares, robust regression analysis and methods of the Committee was used.

Using developed during the this work criteria bursts of particles were selected which may be associated with local precipitations of electrons from the inner radiation belt according ARINA experiment.

The results of the analysis of the burst registered on August 24, 2009 in the course of ARINA experiment. For this the burst geographical location of precipitation electrons  from the inner radiation belt was found. The precipitation area includes Mid-Atlantic ridge, which is an active seismic area.

\bigskip % extra skip inserted
\begin{acknowledgments}	
	This  work  was  supported  by  National  Research Nuclear  University  MEPhI  in  the  framework  of  the Russian  Academic  Excellence  Project  (contract  No. 02.a03.21.0005, 27.08.2013).
\end{acknowledgments}

\bigskip % extra skip inserted

\begin{thebibliography}{9}   % Use for  1-9  references
%\begin{thebibliography}{99} % Use for 10-99 references

\bibitem{burstcite}
S.~Yu.~Aleksandrin et al., ``Interrelation between energy and time distributions of high-energy electrons during the observation of the particle bursts in the near-Earth space'', Journal of Physics: Conference Series, ECRS-2014, 2014

\bibitem{arinaexp}
A.~V.~Bakaldin et al., ``Satellite experiment ARINA for studying seismic effects in high-energy particle fluxes in Earth's magnetosphere'', Cosmic Research, 2007 (In~Russian)

\bibitem{icrc2015}
T.~R.~Zharaspayev et al., ``Robust regression analysis of energy spectrum evolution in time for relativistic electron bursts in Earth’s magnetosphere'', Proceedings of science, 2015, POS(ICRC2015)096

\bibitem{zhou}
Z.~H.~Zhou, ``Ensemble Methods: Foundations and Algorithms.'' -- CRC Press, 2012.

\end{thebibliography}

\end{document}